\documentclass[twocolumn,prl,showpacs]{revtex4}

\usepackage{graphicx}% Include figure files
\usepackage{dcolumn}% Align table columns on decimal point
\usepackage{bm}% bold math

\newcommand{\be}{\begin{equation}}
\newcommand{\ee}{\end{equation}}
\newcommand{\ba}{\begin{eqnarray}}
\newcommand{\ea}{\end{eqnarray}}

\begin{document}
\title{Cosmic Radiation Constraints on Low String Scale and Extra Dimension
Cross Sections}
\author{G{\"u}nter Sigl}
\affiliation{GReCO, Institut d'Astrophysique de Paris, C.N.R.S.,
98 bis boulevard Arago, F-75014 Paris, France}

\begin{abstract}
The observed interaction energy of cosmic rays with atmospheric nuclei
reaches up to a PeV in the center of mass. We compute
nucleon-nucleon and nucleon-neutrino cross sections for various
generic parton cross sections appearing in string and brane world
scenarios for gravity and compare them with cosmic ray data. Scenarios with
effective energy scales in the TeV range and parton cross sections
with linear or stronger growth with the center of mass energy appear
strongly constrained or ruled out. String-inspired scenarios with
infinite-volume extra dimensions may require a fundamental scale
above $\simeq100\,$eV for which they
are probably in conflict with gravity on parsec scales.
\end{abstract}


\maketitle

{\it Introduction:}
The idea that new physics will appear at an energy of $\simeq1\,$TeV
is currently very popular. Many of the scenarios discussed in the
literature involve extra dimensions and aim at unifying gravity
with the Standard Model interactions, see, e.g. Ref.~\cite{add}.
Scenarios with a quantum gravity/string scale as small as in the
sub-eV range have been proposed to explain the smallness of the
observed cosmological constant in consistency with collider
experiments, cosmology, and gravity measurements~\cite{dgln,dgs}. 

On the other hand cosmic ray interactions have been observed up
to a PeV in the center of mass (CM)~\cite{uhecr}, about a factor
thousand higher than reached in accelerator experiments. Although rather
indirect, these observations are consistent with interactions
extrapolated from their Standard Model description and thus suggest
the absence of dramatic new effects at a TeV. Since new interactions
above a TeV are usually weak, this does not necessarily put strong
constraints on new physics. However, there are two effects that
can act as ``magnifiers'' of weak individual interactions: First,
when probed at very high energies, the nucleons appear to consist
of partons each of which acts as a target for new interactions
and whose numbers grow with energy roughly as $E^{0.4}$. Second,
theories in which the fundamental constituents are not described
as point particles but as extended objects such as strings, often
predict numbers of excitable states that increase as power laws
or even exponentially with energy.

In the present paper we parametrize new parton level cross
sections by a mass scale $M_{\rm eff}$, an integer $n$ characterizing
the spin of possible new states produced in the interactions,
and a maximal squared 4-momentum transfer $t_{\rm max}$.
We then establish constraints on these parameters from data
on ultra-high energy (UHE) cosmic rays (CR) and neutrinos.

{\it New Fundamental Cross Sections:}
We parametrize the fundamental cross section between Standard Model
elementary particles of type $i$ and $j$ involving other Standard
Model particles and at least one bulk state by
\begin{equation}
  \sigma_{ij}(s)\sim\frac{t_{\rm max}}{s^2}
  \left(\frac{s}{M_{\rm eff}^2}\right)^{1+n/2}
  \,.\label{sigma_parton}
\end{equation}
Here, $n$ is a constant, $M_{\rm eff}$ is the effective mass scale, and
$s$ and $t$ with $|t|\leq t_{\rm max}$ are the usual Mandelstam variables.

There are several scenarios for which expressions such
as Eq.~(\ref{sigma_parton}) appear. In the case of $n$ large compact
extra dimensions~\cite{add} Kaluza-Klein (KK) gravitons can be produced
with cross sections $\sigma_g\sim M_{\rm Pl}^{-2}$, where
$M_{\rm Pl}\simeq1.72\times10^{18}\,$GeV is the reduced
Planck mass~\cite{grw}. For flat extra dimensions the total number of KK
states is $\sim s^{n/2}V_n$,
where $V_n$ is the volume of the extra dimension. Using the
relation $M_{\rm Pl}^2=M_{\rm eff}^{n+2}V_n$, this leads to
Eq.~(\ref{sigma_parton}) with $n$ interpreted as the number of
extra dimensions and $M_{\rm eff}$ the $4+n$ dimensional fundamental
gravity scale. Above $M_{\rm eff}$ the growth of the cross section
Eq.~(\ref{sigma_parton}) for gravitational scattering is softened
by unitarity effects~\cite{eikonal} within scenarios involving point
particles. To obtain conservatively low cross sections we will
assume an exponential cut-off $\exp(-s^{1/2}/M_{\rm eff})$ in such
scenarios, whereas $t_{\rm max}\simeq s$.

It has been realized that in extra dimension scenarios the
largest contribution to the cross section for collisions above
$M_{\rm eff}$ may be the production of $3+n$ dimensional black
holes~\cite{bh}.
The corresponding cross sections have been estimated by the
geometric cross section $\pi r_s(s)^2$ defined by the Schwarzschild
radius in $3+n$ spatial dimensions $r_s(s)\propto s^{\frac{1}{2(1+n)}}$.
This is also of the form Eq.~(\ref{sigma_parton}) with $n$ substituted by
$2/(1+n)$. It has further been argued that semi-classical p-branes
completely wrapped around $p$ small extra dimensions leads to
cross sections with the scaling $s^{\frac{1}{(1+n)}}$ where $n$ is now
interpreted as the number of large extra dimensions around which
the p-brane is not wrapped~\cite{aco}.

Recently scenarios have been discussed where the fundamental quantum gravity
scale is as low as $M_{\rm s}\sim10^{-3}\,$eV~\cite{dgln}. Such
scenarios can be realized, for example, if the Standard Model lives
on a 3-brane and is coupled to gravity propagating in 5 or more
infinite-volume space-time dimensions with a fundamental scale
$M_{\rm s}$. Thus,
gravity is strong in the bulk but non-gravitational interactions
on our 3-brane induce the observed Planck scale $M_{\rm Pl}$ and
thus shield strong gravity from the bulk. Above the fundamental
scale $M_{\rm s}$ gravity is assumed to be regularized
and thus has its effective coupling unaffected~\cite{dgln}. This
is the case in string theory where $M_{\rm s}$ is identified with
the string scale and excitations of mass $m^2=4(N-1)M_{\rm s}^2$
appear at integer levels $N\geq1$~\cite{polchinski}
with spin $j\leq2N$ (for closed strings). Interactions of Standard
Model particles on
the brane can then lead to stringy bulk states of spin $j$ with
$2\leq j\lesssim s/M_{\rm s}^2$. The cross section for each of these
states is $\sigma\sim(t_{\rm max}/s^2)(s/M_{\rm Pl}^2)^{j-1}$
[see Eq.~(6.6) in Ref.~\cite{dgln}], but the number of states grows
as $(s/M_{\rm s}^2)^{j-1}$. Thus, the intermediate scale in
Eq.~(\ref{sigma_parton}) arises as
\begin{equation}
  M_{\rm eff}=(M_{\rm Pl}M_{\rm s})^{1/2}\,,\label{meff}
\end{equation}
with $n=4j-6$, and this scaling reflects the usual Regge behavior. To
recover the correct graviton zero-mode coupling we have to use
$t_{\rm max}=s$ for $n=j=2$ in Eq.~(\ref{sigma_parton}).
In the string theory context the scaling Eq.~(\ref{sigma_parton}) is not
expected to be strongly modified above $M_{\rm eff}$. The usual
unitarity bound for point particles, $\sigma_{ij}\lesssim s^{-1}$,
does not apply to extended strings.

{\it Nucleon-Nucleon and Neutrino-Nucleon Cross Sections:}
Total cross sections are obtained by folding Eq.~(\ref{sigma_parton})
with the parton distributions $f_i(x,Q)$ in the nucleon. We neglect factors
from the spin structure and follow an approach similar to
Ref.~\cite{rt} leading to
\begin{eqnarray}
  \sigma_{NN}(s)&\sim&\sum_{ij}\int_0^1 dx_1dx_2 f_i(x_1,Q)f_j(x_2,Q)
  \sigma_{ij}(x_1x_2s)\,,\nonumber\\
  \sigma_{\nu N}(s)&\sim&\sum_i\int_0^1 dx f_i(x,Q)\sigma_{\nu i}(xs)
  \,.\label{sigma}
\end{eqnarray}
Here, the sums run over gluons and quarks. Note that in principle
there is another integral $\int_0^{\hat{s}}dQ^2(d\sigma_{ij}
(\hat{s},Q^2)/dQ^2)\cdots$ over the four-dimensional energy-momentum
transfer $Q^2=-t$,
where $\hat{s}=x_1x_2s$ or $\hat{s}=xs$, respectively. We assume
$t_{\rm max}\simeq s$ in Eq.~(\ref{sigma_parton}) and have
approximated this integral by the total parton cross section
$\sigma_{ij}(\hat{s})$ multiplied by the parton distributions taken
at $Q^2=\hat{s}$ which is justified within an order of magnitude
estimate in the absence of concrete models. For the parton distributions
we use the CTEQ6 distributions in electronic form from Ref.~\cite{cteq}.

\begin{figure}[ht]
\includegraphics[width=0.48\textwidth,clip=true]{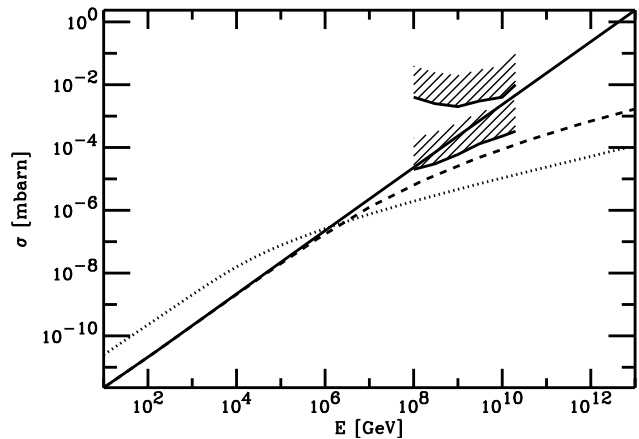}
\caption[...]{The neutrino-nucleon neutral current cross section
$\sigma_{\nu N}$ from Eq.~(\ref{sigma}) involving spin-2
emission into the bulk [$n=2$, $t_{\rm max}\simeq s$ in
Eq.~(\ref{sigma_parton})],
for $M_{\rm eff}=1.3\,$TeV with (dashed line) and without (solid line)
an exponential cut-off at $M_{\rm eff}$. In the string scenario
this corresponds to $M_{\rm s}=10^{-3}\,$eV. The Standard Model
neutral current cross section is shown as the dotted line for
comparison. The two hatched areas indicate cross sections excluded
by air shower data considerations, assuming that a considerable
part of the final products are visible (see text).}
\label{F1}
\end{figure}

Fig.~\ref{F1} shows the neutrino-nucleon neutral current cross section
for spin-2 emission into the bulk obtained for the two scenarios.
Here, $s=2Em_N$, where $E$ is the incoming neutrino energy and
$m_N$ is the nucleon mass.
We stress that in the absence of a concrete model the uncertainties
due to the exact form of the coupling, the spin structure etc. can
be of the order 100. Also the observable signatures of such
reactions depend on the energy fraction $f_v$ going into
visible Standard Model particles and thus on currently unknown
details. Assuming, for example, that observable particle production
is cut off at $Q^2\gtrsim M_{\rm eff}^2$ where most of the energy
goes into invisible bulk states, results in $f_v\sim0.1$ for a
$10^{19}\,$eV neutrino interacting with a nucleon.
Nevertheless the results in Fig.~\ref{F1} provide qualitative indications.
Note that the cross section can be scaled by $M_{\rm eff}^{-2}$ for
different effective mass scales.

{\it Phenomenological Consequences:}
Fig.~\ref{F1} implies neutrino-nucleon cross sections significantly
higher than in the Standard Model at energies above $\simeq1\,$PeV
which would make neutrinos easier to detect. Specifically, cross sections
$\sigma_{\nu N}\lesssim1\,$mb would give rise to deeply penetrating
atmospheric air showers. Neutrinos are expected to be produced by
UHECR which produce pions on the cosmic microwave background every few
Mpc above $\simeq4\times10^{19}\,$eV~\cite{gzk}. The observation of such
energetic UHECR which are believed to have an extragalactic origin~\cite{bs},
allows to estimate the resulting secondary flux of ``cosmogenic''
neutrinos between $\simeq10^{17}\,$eV and $\simeq10^{19}\,$eV within
a factor $\simeq100$~\cite{stecker,nu_fluxes,kkss}. For a differential
neutrino flux $\phi_\nu(E)$,
a detector whose sensitivity to deeply penetrating showers of energy $E$
effectively corresponds to $N(E)$ target nucleons would see a rate
of such showers per solid angle given by
$R(E)\simeq\phi_\nu(E)\sigma_{\nu N}(E)N(f_vE)$. No such showers consistent
with weakly interacting primaries have been observed yet which,
for a given cosmogenic flux $\phi_\nu(E)$, lead to upper limits
on $\sigma_{\nu N}(E)$~\cite{mr,tol,rt}. Fig.~\ref{F1} shows the
resulting excluded range of cross sections, assuming $f_v\sim1$,
for a conservative and an optimistic neutrino flux estimate~\cite{tol,rt}.
The optimistic flux assumes that the neutrino energy fluence is
comparable to the isotropic $\gamma-$ray energy fluence in the
Universe~\cite{kkss}.

Thus, cross sections that are larger than the Standard Model cross section by
a factor 10-1000 and smaller than $\sim1\,$mb for
$10^{17}\,{\rm eV}\lesssim E\lesssim10^{19}\,$eV would imply
neutrino fluxes smaller than expected.
The projected sensitivity of future experiments~\cite{kkss}
indicate that these limits could be lowered down to the Standard Model
cross section. We conclude that parton cross sections with
$M_{\rm eff}\lesssim1\,$TeV, $t_{\rm max}\simeq s$,
and $n\gtrsim2$ in Eq.~(\ref{sigma_parton}) are most likely in conflict
with the non-observation of deeply penetrating air showers, especially
in the stringy context where no strong unitarization cut-off
above $M_{\rm eff}$ is expected. In this latter scenario, Eq.~(\ref{meff})
implies the limit $M_{\rm s}\gtrsim10^{-3}\,$eV independent of similar
limits from other considerations~\cite{dgln}.

In the string context the $j=n=2$ (graviton) contribution considered
above in Fig.~\ref{F1} provides a lower limit to the total cross section. The
contribution of spin $j\gg2$ state production leads to a threshold
behavior: The behavior of the
Veneziano amplitude in the hard scattering limit, $s,t\gg M_{\rm s}^2$,
$t/s=$const., implies that individual amplitudes for $j\gg1$
are usually suppressed by $\sim\exp[-|t|/M_{\rm s}^2]$~\cite{polchinski},
and thus $t_{\rm max}\simeq M_{\rm s}^2$ in Eq.~(\ref{sigma_parton}).
Now summing over all
$j\lesssim s/M_{\rm s}^2$ results in an exponential growth
of the form $\sigma_{ij}(s)\sim s^{-1}\exp\left[(s^{1/2}-M_{\rm eff})/
M_{\rm s}\right]$~\cite{kovesi-domokos}, and similarly for cross sections
involving nucleons. This leads to Hagedorn-like saturation
at $s\simeq M_{\rm eff}^2$, both at the parton level and after
folding in Eq.~(\ref{sigma}). However, air showers with energies
up to a few times $10^{20}\,$eV, or $s\simeq(1\,{\rm PeV})^2$ have been
observed~\cite{uhecr}. The shape and starting point in the atmosphere
of these showers are consistent with nucleon primaries and cross
sections with atmospheric nuclei that are within
about a factor 2 of the expected Standard Model hadronic cross sections.
Therefore, models with $M_{\rm s}\ll M_{\rm eff}$ and
$M_{\rm eff}\lesssim1\,$PeV should be ruled out. Furthermore,
observation of an air shower of energy $E$ would require a primary
of energy $E/f_v$. If visible particles are only produced for
$Q^2\lesssim M_{\rm eff}^2$, for $M_{\rm eff}\sim1\,$TeV Eq.~(\ref{sigma})
leads to $f_v\lesssim10^{-3}$ for an incoming nucleon energy
$E\gtrsim10^{20}\,$eV, and thus much higher energy primaries
would be necessary to explain the observations.

This has important phenomenological ramifications for scenarios with
a quantum gravity/string scale in the sub-eV range which has recently been
proposed to explain the smallness of the observed cosmological
constant~\cite{dgs}. There, the above argument together with
Eq.~(\ref{meff}) implies $M_{\rm s}\gtrsim100\,$eV.
In these scenarios Newtonian gravity is modified on
scales $r\lesssim M_{\rm s}^{-1}\simeq2\,\mu{\rm m}\,
(M_{\rm s}/100{\rm eV})^{-1}$, where massive graviton exchange
becomes relevant. This is consistent with gravity measurements on
sub-mm scales~\cite{submm}. At very small momenta
$p\lesssim M_{\rm s}^2/M_{\rm Pl}$, the propagator for the
zero-mode graviton in four dimensions is modified to
$\simeq(M_{\rm Pl}^2p^2+aM_{\rm s}^4)^{-1}$~\cite{dghs} in case of at
least two infinite-volume extra dimensions, where $a$ is
a constant of order unity. This makes gravity appear higher-dimensional,
i.e. decreasing with a higher power of distance than Newtonian gravity,
on scales $r\gtrsim M_{\rm Pl}/M_{\rm s}^2\simeq1\,{\rm pc}
\,(M_{\rm s}/100{\rm eV})^{-2}$~\cite{dvali}. The above bound would thus imply
modifications of gravity on parsec scales which seems
phenomenologically unviable. This problem can be avoided for only
one extra dimension where gravity is modified only at much larger
scales $r\gtrsim M_{\rm Pl}^2/M_{\rm s}^3$~\cite{dghs}, however, this does
not allow to explain the smallness of the cosmological constant~\cite{dgs}.
An unlikely loop-hole consists of UHECR primaries consisting of
$\gtrsim10^3$ nucleons.

{\it Conclusions:}
Parton cross sections involving an effective mass scale
$M_{\rm eff}\lesssim1\,$TeV and growing linearly or faster in
the squared CM energy are likely in conflict
with the non-observation of deeply penetrating air showers.
Extra-dimensional black hole or p-brane production cross
sections increase more slowly than $s$ and are thus not
strongly constrained yet above a TeV, consistent with findings
in the literature~\cite{rt}.
We further argued that in scenarios with a Hagedorn-like saturation
at $M_{\rm eff}$ due to an exponential increase of states,
this mass scale should be larger than the largest CM energies
observed in interactions with hadronic or smaller cross sections,
which is currently about a PeV.

In possible string theoretic descriptions of quantum gravity with a
scale $M_{\rm s}\ll1\,$TeV, the lower limit on $M_{\rm eff}$ translates
into $M_{\rm s}\gtrsim100\,$eV via Eq.~(\ref{meff}). This bound
could further increase proportional to the maximal UHECR energy
that may be seen with much bigger experiments now under construction
such as the Pierre Auger project~\cite{auger}.

In scenarios which also explain the smallness of the cosmological constant
with more than two infinite-volume extra dimensions the lower
bound on $M_{\rm s}$ would lead to modifications of Newtonian gravity
on scales $r\gtrsim1\,{\rm pc}\,(M_{\rm s}/100{\rm eV})^{-2}$. Such scenarios
thus appear to be in conflict either with UHECR observations or
with gravity on parsec scales.

{\it Acknowledgements:}
We thank Gia Dvali, David Langlois, Karim Malik, and Lorenzo Sorbo
for illuminating discussions.

%%%%%%%%%%%%%%%%%%%%%%%%%%%%%%%%%%%%%%%%%%%%%%%%%%%%%%%
%%%%%%%%%%%%%%%%%%%%%%%%%%%%%%%%%%%%%%%%%%%%%%%%%%%%%%%%%%%%%%%%%%


\begin{thebibliography}{9}

\bibitem{add} N.~Arkani-Hamed, S.~Dimopoulos, and
G.~Dvali, Phys.~Lett. B 429 (1998) 263; I.~Antoniadis,
N.~Arkani-Hamed, S.~Dimopoulos, and G.~Dvali, Phys.~Lett. B
436 (1998) 257; N.~Arkani-Hamed, S.~Dimopoulos, and G.~Dvali,
Phys.~Rev.~D 59 (1999) 086004.

\bibitem{dgln} G.~Dvali, G.~Gabadadze, M.~Kolanovi\'{c}, and
F.~Nitti, Phys.~Rev.~D 65 (2001) 024031.

\bibitem{dgs} G.~Dvali, G.~Gabadadze, and M.~Shifman,
e-print hep-th/0202174.

\bibitem{uhecr} for a recent review of the experimental
situation, see, e.g., M.~Nagano and A.~A.~Watson, Rev.~Mod.~Phys.
72 (2000) 689.

\bibitem{grw} G.~F.~Giudice, R.~Rattazzi, and G.~D.~Wells,
Nucl.~Phys. B544 (1999) 3.

\bibitem{eikonal} see, e.g., P.~Jain, D.~W.~McKay, S.~Panda, and
J.~P.~Ralston, Phys. Lett. B484 (2000) 267;
R.~Emparan, M.~Masip, and R.~Rattazzi, Phys.~Rev.~D 65 (2002) 064023.

\bibitem{bh} see, e.g., T.~Banks and W.~Fischler, e-print hep-th/9906038;
S.~Dimopoulos and R.~Emparan, Phys.~Lett. B526 (2002) 393.

\bibitem{aco} E.-J.~Ahn, M.~Cavaglia, and A.~V. Olinto,
e-print hep-th/0201042.

\bibitem{polchinski} see, e.g., J.~Polchinski, {\it String Theory}
(Cambridge University Press, Cambridge, England, 1998) Vol.~1.

\bibitem{rt} A.~Ringwald and H.~Tu, Phys.~Lett.~B 525 (2002) 135.

\bibitem{cteq} J.~Pumplin et al., e-print hep-ph/0201195.

\bibitem{gzk} K.~Greisen, Phys.~Rev.~Lett. 16 (1966)
748; G.~T.~Zatsepin and V.~A.~Kuzmin, Pis'ma
Zh. Eksp. Teor. Fiz. 4 (1966) 114 [JETP. Lett. 4 (1966) 78].

\bibitem{bs} for a review see, e.g., P.~Bhattacharjee and G.~Sigl,
Phys.~Rept. 327 (2000) 109.

\bibitem{stecker} F.~W.~Stecker, Astrophys.~J. 228 (1979) 919.

\bibitem{nu_fluxes} S.~Yoshida and M.~Teshima, Prog.~Theor.~Phys.
89 (1993) 833; R.~J.~Protheroe and P.~A.~Johnson,
Astropart.~Phys. 4 (1996) 253; S.~Yoshida, H.~Dai, C.~C.~Jui,
and P.~Sommers, Astrophys.~J. 479 (1997) 547; R.~Engel and
T.~Stanev, Phys.~Rev.~D 64 (2001) 093010.

\bibitem{kkss} for a recent detailed discussion see, e.g., O.~E.~Kalashev,
V.~A.~Kuzmin, D.~V.~Semikoz, and G.~Sigl, e-print hep-ph/0205050.

\bibitem{mr} D.~A.~Morris and A.~Ringwald, Astropart.~Phys. 2 (1994)
43.

\bibitem{tol} C.~Tyler, A.~V.~Olinto, and G.~Sigl,
Phys.~Rev.~D 63 (2001) 055001.

\bibitem{kovesi-domokos} G.~Domokos and S.~Kovesi-Domokos,
Phys.~Rev.~Lett. 82 (1999) 1366.

\bibitem{submm} C.~D.~Hoyle et al., Phys.~Rev.~Lett. 86 (2001) 1418.

\bibitem{dghs} G.~Dvali, G.~Gabadadze, X.~Hou, and E.~Sefusatti,
e-print hep-th/0111266.

\bibitem{dvali} I am grateful to G.~Dvali for pointing this out.

\bibitem{auger} J.~W.~Cronin, Nucl.~Phys.~B (Proc.~Suppl.) 28B (1992)
213; The Pierre Auger Observatory Design Report (ed.~2), March 1997;
see also {\sf http://www.auger.org}.

\end{thebibliography}
\end{document}